\documentclass{ws-ijmpc}
\usepackage[super]{cite}
\usepackage{xcolor}
\usepackage[verbose,hypertexnames=false]{hyperref}
\hypersetup{colorlinks=false,allbordercolors=blue,pdfborderstyle={/S/U/W 1}}

\begin{document}

\markboth{E. M. K. Souza \& R. N. C. C. Leite \& A. M. C. Souza}
{Chesapeake Bay Food Web: Robustness Analysis via Energy Cutoff in Complex Networks}

\catchline{}{}{}{}{}

\title{Chesapeake Bay Food Web: Robustness Analysis via Energy Cutoff in Complex Networks}

\author{Eduardo M. K. Souza}

\address{Instituto de Fisica, Universidade Federal de Alagoas\\
Maceio AL, 57072-900, Brazil\\
eduardo.souza@fis.ufal.br}

\author{Rafael N. C. C. Leite}

\address{Departamento de Fisica, Universidade Federal de Pernambuco\\
Recife PE, 50670-901, Brazil\\
rafael.nathan@ufpe.br}

\author{Andre M. C. Souza}

\address{Departamento de Fisica, Universidade Federal de Sergipe\\
Sao Cristovao SE, 49100-000, Brazil\\
amcsouza@academico.ufs.br}

\maketitle

\begin{history}
\received{(Day Month Year)}
\revised{(Day Month Year)}
\accepted{(Day Month Year)}
\published{(Day Month Year)}
\end{history}

\begin{abstract}
The Chesapeake Bay, one of the largest estuaries in the United States, is an ecological system of great complexity and relevance. The food web is composed of thirty-six trophic components, all of which are functionally connected. In this work, the interactions among these components are numerically analyzed using complex network methods. An energy flow cutoff paradigm is applied to a weighted ecological network. The results reveal patterns characteristic of connectivity dynamics, evidencing both the initial robustness of the system and its tendency to fragmentation at higher values of the cutoff. From an applied perspective, the findings underscore the importance of conservation strategies that protect keystone species, such as carnivorous fish, which act as crucial connectors between the two main subnetworks. Although they are positioned at the top of the food web and are often assumed to be less critical to network stability, these species play a pivotal role in regulating populations of lower-level organisms, thereby maintaining the overall integrity of the ecosystem.
\end{abstract}


\ccode{PACS number: 89.75.Fb; 87.23.Cc; 87.18.Vf}

\section{Introduction}

Natural ecosystems are highly interconnected systems, composed of a diversity of species that interact with each other through trophic relationships such as predation, herbivory, and decomposition \cite{Ulanowicz1986}. These interactions are organized into complex food chains and webs, and they determine the flows of energy and matter that sustain the ecological dynamics and stability of these systems \cite{Pimm1984}. Historically, these food chains have been represented in a simplified linear fashion, but recent approaches have shown that the complexity of these interactions requires more realistic models to understand them \cite{Dunne2002EL,Williams2011,Romanuk2019,Liu2025,Sun2021}.

Complex network theory emerged as an interdisciplinary approach capable of representing, quantifying, and analyzing systems composed of multiple interconnected elements \cite{Newman2010}. Originally, network theory was developed in physics and mathematics to study systems such as the Internet, social networks, and transportation networks; it has also been increasingly applied in ecology to model trophic, mutualistic, and dispersal interactions. This approach allows for the quantitative analysis of structural properties such as connectivity, robustness, centrality, and hierarchy within ecosystems \cite{Boccaletti2006,Estrada2007,MontoyaSole2002}.

By applying network theory to ecological contexts, we gain a more comprehensive perspective on how species are functionally connected, how disturbances can propagate through the network, and what are the key elements that maintain the integrity of the system \cite{Callaway2000,Albert2000}. This is especially relevant in times of ecological crisis, where loss of biodiversity, climate change, and anthropogenic impacts threaten the structure and functionality of natural ecosystems \cite{Sole2001,AllesinaTang2012}.

In this context, the work of Baird and Ulanowicz \cite{BairdUlanowicz1989} provides a solid foundation by quantifying energy flows among thirty-six trophic components of the Chesapeake Bay, one of the largest estuaries in the United States and an ecological system of great complexity and relevance. Their approach stands out for quantifying energy transfers between the living and nonliving elements of the ecosystem, offering a detailed representation of trophic relationships.

Inspired by this work, as well as by the modern approach of Dunne \cite{Dunne2002PNAS,Dunne2002EL}, who analyzed multiple food webs using complex network methods, we present a numerical analysis of the Chesapeake Bay food web, focusing exclusively on its living components. Therefore, our goal is to understand how variations in the interaction threshold (cutoff) affect the food web structure and how this can offer new insights about resilience, structure vulnerabilities, and the functional importance of ecosystem components.

In what follows, Sec. \ref{Metodologia}, we describe the methodology used to construct the trophic networks and the topological metrics employed. In Sec. \ref{Resultados}, we present our results, including the structural degradation of the network and the ecological percolation transition. In Sec. \ref{Discussao}, we discuss the implications for ecological resilience and network theory. Finally, our main conclusions and perspectives are summarized in Sec. \ref{Conclusoes}.

\section{Methodology} \label{Metodologia}

Our starting point is the energy transfer matrix originally published by Baird and Ulanowicz \cite{BairdUlanowicz1989}, which describes the energy fluxes between thirty-six trophic components of the Chesapeake Bay. In this study, we considered only the thirty-three living components of the ecosystem, discarding the non-living ones. Each component (site) represents a species or functional group within the food web: (1) Phytoplankton; (2) Attached bacteria; (3) Sediment bacteria; (4) Benthic diatoms; (5–7) Microzooplankton; (8–10) Zooplankton; (11–13) Suspension feeders; (14–19) Deposit feeders; (20–24) Suspension-feeding fish; (25–29) Benthic-feeding fish; (30–33) Carnivorous fish.

Each element $A_{ij}$ of the original matrix represents the energy transfer rate from component \textit{i} to component \textit{j}. To transform this flow matrix into a symmetric binary adjacency matrix, we introduce a threshold parameter called cutoff $\theta\in[0,100]$. For each cutoff value between 0 and 100, elements $A^{(\theta)}_{ij}$ greater than the cutoff value imply the existence of a connection between nodes \textit{i} and \textit{j}, so that $A^{(\theta)}_{ij}=1$. Otherwise, $A^{(\theta)}_{ij}=0$. Thus, we construct a series of symmetric networks corresponding to different levels of filtering of the energy interactions.

From these networks, we calculate different topological properties of network theory \cite{Newman2010}: (i) Average degree $\langle k\rangle$: average number of edges per node; (ii) Connectance $C=2E/[N(N-1)]$: proportion of existing edges in relation to the maximum possible number of edges; (iii) Clustering coefficient: measure of the tendency to form triangles or local cohesive groups; (iv) Average distance $L$: average shortest path between all pairs of nodes; (v) Number of subnetworks $S$ and number of connected components in the largest subnetwork $G$.  

The analysis is structured in complementary stages to investigate different aspects of the system: (i) Study of structural parameters as a function of the cutoff $\theta$: we evaluate how topological metrics vary when the $\theta$ is progressively increased from 0 to 40, the region in which the network remains connected in a single cluster; (ii) Network robustness by site removal via $\theta$: we analyze how the progressive elimination of connections (increasing the $\theta$) leads to fragmentation and the weakening of global connectivity, identifying critical transition points; (iii) Assessment of fragmentation dynamics: we study the evolution of the number of subnetworks and the size of the largest connected component across the entire $\theta$ range (0 to 100), observing when the network ceases to be globally connected; (iv) Dynamic visualization: we use a video \cite{video} of the evolution of the network with increasing $\theta$ to interpret substructure formation, cluster persistence, and the relative importance of sites in maintaining cohesion.

This integrated approach allowed us to understand not only the static structure of the network but also its dynamic behavior and its resilience to the gradual removal of trophic interactions.

\section{Results} \label{Resultados}

The analysis of the structural properties of the Chesapeake Bay ecological network as a function of the cutoff $\theta$ reveal patterns characteristic of connectivity dynamics in complex networks, evidencing both the initial robustness of the system and its tendency to fragmentation at higher values of the parameter.

\begin{figure}[b]
\centerline{\includegraphics[width=11cm]{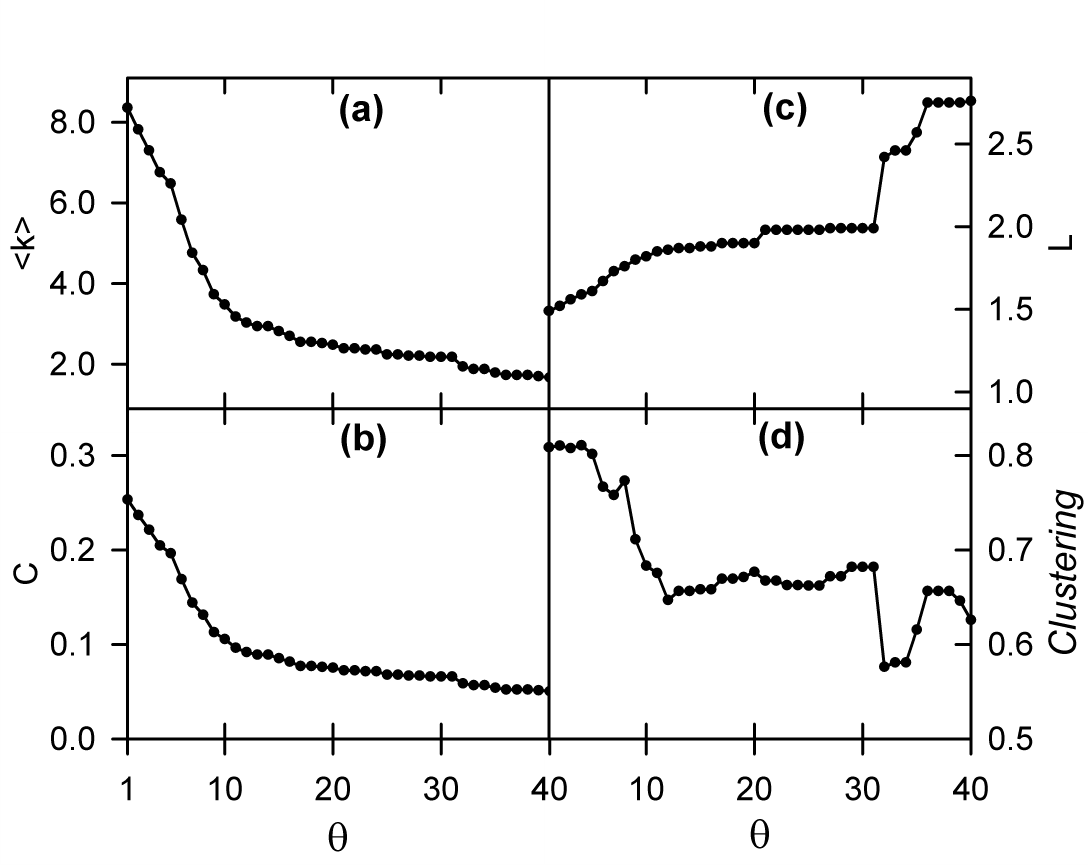}}
\vspace*{8pt}
	\caption{(a) Average degree $\langle k\rangle$, (b) Connectance $C$, (c) Average distance $L$, (d) \emph{clustering} as a function of $\theta$ for the globally connected network regime ($0\le\theta\le40$).}
	\label{fig1}
\end{figure}

As we can see in Fig. \ref{fig1}, we report $\langle k\rangle$, $C$, $L$ and \emph{clustering} for the globally connected network regime ($0\le\theta\le40$). In Fig. 1(a), a sharp decrease in the average degree is observed, going from values close to $\langle k\rangle = 8$  to approximately $\langle k\rangle = 2$ when $\theta$ reaches 40. This behavior indicates that the network, initially dense in connections due to the inclusion of all energetic interactions (even the weakest ones), progressively loses secondary links as $\theta$ increases. In ecological terms, this represents the filtering of energetically unimportant interactions, leaving only the most intense energy flows and thereby reducing network redundancy.

In parallel, we present the connectance in Fig. 1(b), which also shows a significant drop from approximately $C = 0.3$ to less than $C = 0.05$. This drop reflects the transition from a highly connected network to a sparse network in which only a small fraction of the possible interactions are maintained. The loss of connectance implies less redundancy of the energy paths, suggesting a reduction in structural resilience to disturbances.

Fig. 1(c) shows that despite the elimination of connections, the average distance $L$ between pairs of nodes increases relatively smoothly. This means that even with fewer interactions, the network maintains a considerable level of efficiency in transporting energy between components. This result indicates the presence of a robust structural core that supports global connectivity, ensuring that relevant energy flows continue to be transmitted efficiently up to a $\theta$ close to 40.  However, the average \emph{clustering} in Fig 1(d) decays more slowly than $\langle k\rangle$ and $C$, remaining relatively high throughout much of the interval. This persistence indicates local redundancy (prey–predator–competitor triangles), which sustains intramedullary cohesion even when global connectivity degrades.

The initial characterization reveals that the network has high relative connectivity, a common characteristic in coastal aquatic ecosystems \cite{BairdUlanowicz1989}. The degree distribution shows the presence of few highly connected species (keystone species or hubs), namely Phytoplankton (1), Sediment bacteria (3) and Zooplankton (8-10), and many species with low degrees of connection, as can be seen in Fig. \ref{fig3}, evidencing a pattern analogous to scale-free networks.

During the removal process, with increasing cutoff $\theta$, we observe that the network maintains its structural connectivity up to a critical cutoff $\theta_{c}=40$, which represents 40$\%$ of the maximum cutoff value. This result confirms the typical robustness of complex systems in the face of undirected disturbances \cite{Albert2000}.

\begin{figure}[!h]
\centerline{\includegraphics[width=11cm]{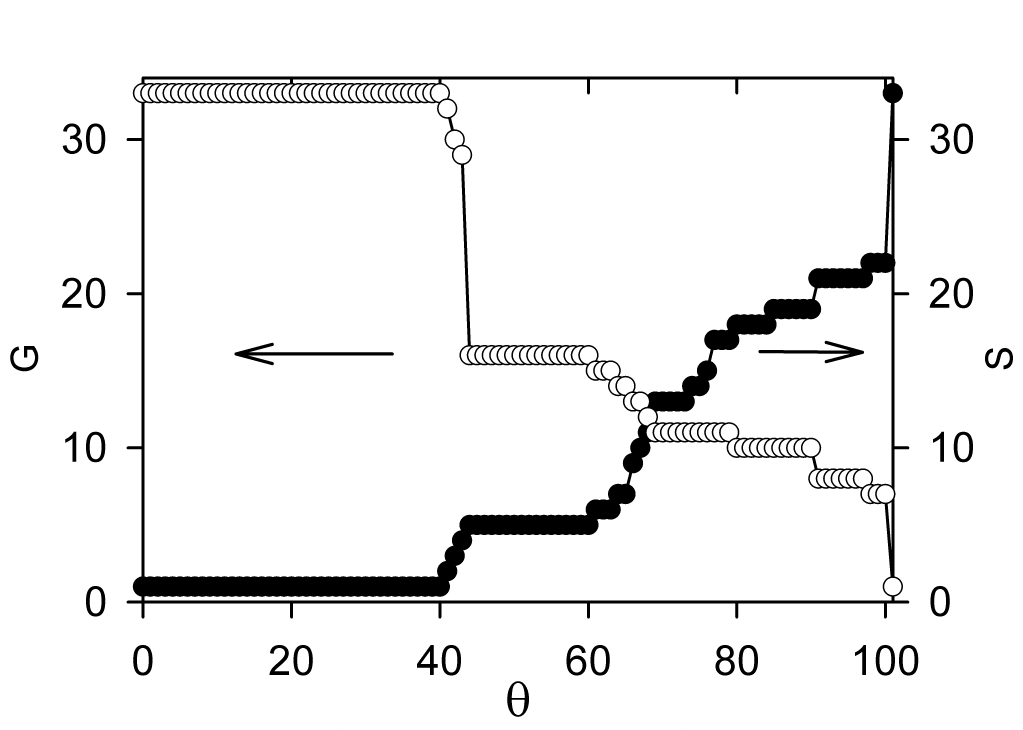}}
\vspace*{8pt}
	\caption{Number of connected components in the largest subnetwork $G$ (left axis) and the number of subnetworks $S$ (right axis) as a function of $\theta$.}
	\label{fig2}
\end{figure}

Fig.~\ref{fig2} covers the entire range of $\theta$, with the number of connected components in the largest subnetwork $G$ on the left axis and the number of subnetworks $S$ on the right axis and shows the evolution of network fragmentation as a function of increasing $\theta$. Also, Fig.~\ref{fig2} reveals a critical point near $\theta_{c}=40$, at which $G$ decays abruptly while $S$ grows rapidly. For $\theta<40$, $G = N$ and $S=1$; just above this threshold, multiple sublattices emerge, characterizing an edge-removal-induced percolation transition \cite{Callaway2000}. At high $\theta$, small clusters (pairs and triads) predominate, $G$ tends to the minimum value ($G=1$), signaling the functional collapse of the system's energy transfer.

To visually illustrate this progressive fragmentation, Fig. \ref{fig3} presents the network structure for four representative cutoff values: (a) $\theta$ = 40, the final point where the network remains connected in a single cluster, where we can already see the presence of three hubs (Phytoplankton (1), Sediment bacteria (3), and Zooplankton (8-10)). Furthermore, it is possible to perceive a network with high clustering. (b) $\theta$ = 55, the first signs of fragmentation are already visible. The network is divided into two large components that connect most sites, but the separation of small groups is already observed (for example, the isolated Attached bacteria (2) and Benthic diatoms (4) sites, and a small set of interconnected microzooplankton 5 and 6). Despite this initial fragmentation, the global connectivity is still maintained. (c) $\theta$ = 69, we reach the point where the number of sites in the largest component of the network is smaller than the number of subnetworks present, as seen in Fig. \ref{fig2}. With this intensification of fragmentation, three main subnetworks can be observed, each formed by semi-functional guilds: One dominated by Deposit feeders, Benthic-feeding fish, and the hub (Sediment bacteria). Another smaller subnetwork formed by Phytoplankton and Suspension feeders, and a third subnetwork with greater diversity containing Zooplankton, Suspension-feeding fish, and Carnivorous fish. We can see that there is no longer a dominant component and that the system loses its ability to maintain energetic cohesion on a global scale. Thus, each functional group ends up isolating itself in a certain subchain formed by two to three groups.

Reaching (d) $\theta$ = 100, the point preceding the complete breakdown of the network. The system is approaching functional collapse. Most sites are found in small groups or isolated. With greater resilience, only the subnetwork formed by Deposit feeder, Benthic fish, and the hub (Sediment bacteria) remains, demonstrating that its a site with high robustness and significant importance to the network.

\begin{figure}[!h]
\centerline{\includegraphics[width=11cm]{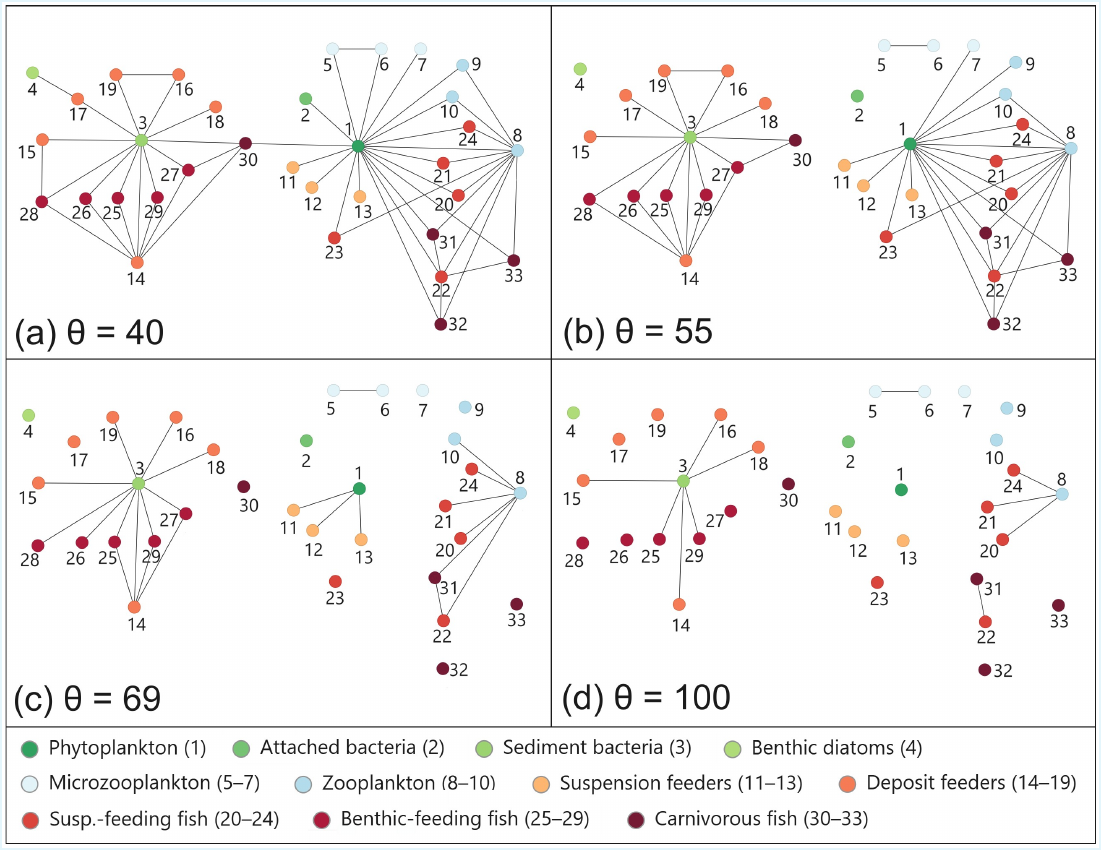}}
\vspace*{8pt}
	\caption{Network structure for four representative cutoff values: (a) $\theta = 40$, (b) $\theta = 55$, (c) $\theta = 69$, (d) $\theta = 100$.}
	\label{fig3}
\end{figure}

Ecologically, this suggests that weak links act as “bridges” between trophic modules. Their removal disconnects guilds that, although internally cohesive (high clustering), depend on few intermodular ties to maintain global cohesion. This pattern is consistent with robustness to random failures and vulnerability to targeted losses of critical interactions \cite{Dunne2002EL, AllesinaTang2012}.

Furthermore, it is observed that the functional centrality of primary producers and consumers (Phytoplankton and Sediment bacteria) is observed at low $\theta$, providing multiple shortest paths. Top predators (Carnivorous fish), although less connected, occupy strategic positions: when they become isolated, they reduce alternative energy routes and accelerate fragmentation at higher levels \cite{Dunne2002PNAS}.

All the results become even clearer when observed in the video \cite{video}, which shows in real time the network disconnection dynamics as the cutoff increases, highlighting the collapse and reorganization patterns described above.

\section{Discussion} \label{Discussao}

The patterns observed in our results are consistent with several theoretical approaches to network theory in ecology and provide empirical support for central hypotheses about resilience and stability in ecosystems.

The fragmentation dynamics identified in cutoffs above forty is reminiscent of a percolation transition, well documented in complex network theory \cite{Albert2000,Callaway2000}. In such a transition, the removal of critical connections or structural elements leads to an abrupt loss of global connectivity. In our case, the cutoff acts as a selective filtering mechanism that progressively deactivates energetic connections. The existence of a critical point ($\theta_{c}=40$), where the network begins to collapse, is a strong indication that the system is supported by a base of intermediate-strength connections.

The reduction in clustering indicates a loss of modularity and local cohesion, a behavior contrary to what is observed in healthy natural ecological networks, which tend to exhibit modularity as a mechanism for containing disturbances \cite{Stouffer2011}. Local modules act as dynamic buffers, preventing disturbances in one part of the network from propagating indefinitely. Their dissolution, as evidenced by the clustering analysis, suggests a reduced capacity to contain ecological shocks.

 In our analysis, the robustness is also reflected in the low average distance values for cutoffs up to 40, even with decreasing average degree and connectance. This suggests that the Chesapeake Bay network maintains high trophic communication efficiency up to a certain threshold, reinforcing the notion that natural food web structure evolves under pressures that maximize energy efficiency and fault tolerance.

Our set of results suggests that the Chesapeake Bay food web is simultaneously efficient and vulnerable. Efficient as it maintains cohesion and low diameter even with significant decreases in connections. Vulnerable, since beyond a critical threshold fragmentation is rapid and irreversible, indicating a strong dependence on low-intensity but structuring interactions.

This balance between robustness and fragility is a common characteristic of complex natural networks \cite{Sole2001}, and its analysis can offer important insights into ecosystem conservation. By identifying the nodes and flows that contribute the most to network connectivity and cohesion, it is possible to develop management and protection strategies that focus not only on dominant species but also on crucial energetic interactions.

From an applied perspective, the results highlight the need for conservation strategies that prioritize the protection of keystone species, such as carnivorous fish (30), which connect the two main subnetworks. Since they are at the top of the food web, one might think that they would be unlikely to be of great importance for the conservation of the network. However, they potentially control the overpopulation of lower-level species and thereby ensure the integrity of the entire ecosystem.

\section{Conclusions} \label{Conclusoes}
Numerical analysis revealed that the Chesapeake Bay living food web exhibits a connectivity transition governed by an energetic cutoff up to $\theta=40$, representing $40\%$ of the
maximum cutoff value, the system maintains a giant component with high clustering and moderate distances; above this threshold the network fragments, the size of the largest component decreases, and the number of subnetworks increases. This topological signature suggests that ecological resilience is critically dependent on a background of weak interactions connecting strongly cohesive modules. Identifying and protecting these links (and the species that sustain them) is as important as preserving the apparent trophic hubs. For example, our results reveal that species such as carnivorous fish (30), which connect the two main subnetworks, are essential to maintaining the integrity of the system, contrary to what might be expected given their position at the top of the food web. These predators play a key role in preventing the overpopulation of lower-level species, thus helping to preserve its ecological balance.

Thus, biodiversity conservation should not be guided solely by the absolute number of species, but also by the maintenance of the critical connections that sustain the ecological network. This interdisciplinary approach, which combines biology, physics, and mathematics, is essential to address the challenges posed by the biodiversity crisis in the Anthropocene.

The cutoff paradigm offers a simple lens for studying connectivity transitions in weighted ecological networks. Natural extensions include: (i) directed analyses (maintaining flow asymmetries), (ii) sequential extinction simulations guided by weighted centralities \cite{Estrada2007}, and (iii) integration with ancestry and development metrics \cite{Ulanowicz1986}.
 
\section*{Acknowledgments}
The authors thank L. K. Souza and M. K. Souza for helpful advice. EMKS and RNCCL benefit from the financial support of Brazilian agency CAPES.

\section*{ORCID}
\noindent Eduardo M. K. Souza - \url{https://orcid.org/0000-0002-0871-7431}

\noindent Rafael N. C. C. Leite - \url{https://orcid.org/0009-0005-0750-1718}

\noindent Andre M. C. Souza - \url{https://orcid.org/0000-0003-3023-6770}


\begin{thebibliography}{0}

\bibitem{Ulanowicz1986} R. E. Ulanowicz, {\it Growth and Development: Ecosystems Phenomenology}
(Springer-Verlag, 1986).
\url{https://doi.org/10.1002/bs.3830330206}.

\bibitem{Pimm1984} S. L. Pimm, {\it Nature} {\bf 307}, 321-326 (1984). \url{https://doi.org/10.1038/307321a0}.

\bibitem{Dunne2002EL} J. A. Dunne, R. J. Williams, N. D. Martinez, {\it Ecology Letters} {\bf 5(4)}, 558-567 (2002a). \url{https://doi.org/10.1046/j.1461-0248.2002.00354.x}.

\bibitem{Williams2011} C. J. Meli\'an, C. Vilas, F. Bald\'o, E. González-Orteg\'on, P. Drake, R. J. Williams, {\it Advances in Ecological Research} {\bf 45}, 225-268 (2011). \url{https://doi.org/10.1016/B978-0-12-386475-8.00006-X}.

\bibitem{Romanuk2019} T. N. Romanuk, A. Binzer, N. Loeuille, W. M. A. Carscallen, N. D. Martinez, {\it Scientific Reports} {\bf 9}, 18242 (2019). \url{https://doi.org/10.1038/s41598-019-54443-0}.

\bibitem{Liu2025} X. Liu, M. Liu, D. Zhao, Y. Sun, {\it Journal of Theoretical Biology} {\bf 613}, 112216 (2025). \url{https://doi.org/10.1016/j.jtbi.2025.112216}.

\bibitem{Sun2021} X. Sun, R. Li, {\it International Journal of Modern Physics C} {\bf 32(07)}, 2150046 (2021). \url{https://doi.org/10.1142/S0129183121500467}.

\bibitem{Newman2010} M. E. J. Newman, {\it Networks: An Introduction}
(Oxford University Press, 2010).
\url{https://doi.org/10.1093/acprof:oso/9780199206650.001.0001}.

\bibitem{Boccaletti2006} S. Boccaletti, V. Latora, Y. Moreno, M. Chavez, D.-U. Hwang, {\it Physics Reports} {\bf 424(4-5)}, 175-308 (2006). \url{https://doi.org/10.1016/j.physrep.2005.10.009}.

\bibitem{Estrada2007} E. Estrada, {\it Ecological Complexity} {\bf 4}, 48-57 (2007). \url{https://doi.org/10.1016/j.ecocom.2007.02.018}.

\bibitem{MontoyaSole2002} J. M. Montoya,  R. V. Sol\'e, {\it Journal of Theoretical Biology} {\bf 214(3)}, 405-412 (2002). \url{https://doi.org/10.1006/jtbi.2001.2460}.

\bibitem{Callaway2000} D. S. Callaway, M. E. J. Newman, S. H. Strogatz, D. J. Watts, {\it Physical Review Letters} {\bf 85(25)}, 5468-5471 (2000). \url{https://doi.org/10.1103/PhysRevLett.85.5468}.

\bibitem{Albert2000} R. Albert, H. Jeong, A.-L. Barab\'asi, {\it Nature} {\bf 406}, 378-382 (2000). \url{https://doi.org/10.1038/35019019}.

\bibitem{Sole2001} R. V. Sol\'e, J. M. Montoya, {\it Proceedings of the Royal Society of London. Series B: Biological Sciences} {\bf 268(1480)}, 2039-2045 (2001). \url{https://doi.org/10.1098/rspb.2001.1767}.

\bibitem{AllesinaTang2012} S. Allesina, S. Tang, {\it Nature} {\bf 483}, 205-208 (2012). \url{https://doi.org/10.1038/nature10832}.
\url{ }.

\bibitem{BairdUlanowicz1989} D. Baird, R. E. Ulanowicz, {\it Ecological Monographs} {\bf 59(4)}, 329-364 (1989). \url{https://doi.org/10.2307/1943071}.

\bibitem{Dunne2002PNAS} J. A. Dunne, R. J. Williams, N. D. Martinez, {\it Proceedings of the National Academy of Sciences} {\bf 99(20)}, 12917-12922 (2002b). \url{https://doi.org/10.1073/pnas.192407699}.

\bibitem{video}
\url{https://youtu.be/pXxHlZD0i1Q}.

\bibitem{Stouffer2011}  D. B. Stouffer, J. Bascompte, {\it Proceedings of the National Academy of Sciences} {\bf 108(9)}, 3648-3652 (2011). \url{https://doi.org/10.1073/pnas.1014353108}.

\end{thebibliography}
\end{document}